\documentclass[sigconf]{acmart}

\usepackage{booktabs} 
\usepackage{algorithm}
\usepackage{algorithmic}

\usepackage{amsmath}
\usepackage{color}
\usepackage{url}
\usepackage{multirow}
\usepackage{bm}
\usepackage{subfigure}
\usepackage{algorithm}
\usepackage{enumerate}
\usepackage{balance}

\DeclareMathOperator{\Tr}{Tr}

\def\BibTeX{{\rm B\kern-.05em{\sc i\kern-.025em b}\kern-.08emT\kern-.1667em\lower.7ex\hbox{E}\kern-.125emX}}
    
\copyrightyear{2019} 
\acmYear{2019} 
\setcopyright{acmcopyright}
\acmConference[SIGIR '19] {42nd International ACM SIGIR Conference on Research and Development in Information Retrieval}{July 21--25, 2019}{Paris, France}
\acmBooktitle{42nd Int'l ACM SIGIR Conference on Research and Development in Information Retrieval (SIGIR '19), July 21-25, 2019, Paris, France}
\acmPrice{15.00}
\acmDOI{10.1145/3331184.3331275}
\acmISBN{978-1-4503-6172-9/19/07}

\begin{document}
\title{Deep Collaborative Discrete Hashing with Semantic-Invariant Structure}

\author{Zijian Wang}
\authornotemark[2] 
\affiliation{%
  \institution{The University of Queensland}
  \city{Brisbane}
  \state{Australia}
}
\email{zijian.wang@uq.net.au}

\author{Zheng Zhang}
\authornotemark[2] 
\authornote{indicates corresponding author; $\dagger$ indicates co-first authors with equal contributions.}
\affiliation{%
  \institution{The University of Queensland}
  \city{Brisbane}
  \state{Australia}
}
\email{darrenzz219@gmail.com}

\author{Yadan Luo}
\affiliation{%
  \institution{The University of Queensland}
  \city{Brisbane}
  \state{Australia}
}
\email{lyadanluol@gmail.com}
\author{Zi Huang}
\affiliation{%
  \institution{The University of Queensland}
  \city{Brisbane}
  \state{Australia}
}
\email{huang@itee.uq.edu.au}

\renewcommand{\shortauthors}{Z. Wang, Z. Zhang, Y. Luo, and Z. Huang}

%
\begin{abstract}
Existing deep hashing approaches fail to fully explore semantic correlations and neglect the effect of linguistic context on visual attention learning, leading to inferior performance. This paper proposes a dual-stream learning framework, dubbed Deep Collaborative Discrete Hashing (DCDH), which constructs a discriminative common discrete space by collaboratively incorporating the shared and individual semantics deduced from visual features and semantic labels. Specifically, the context-aware representations are generated by employing the outer product of visual embeddings and semantic encodings. Moreover, we reconstruct the labels and introduce the focal loss to take advantage of frequent and rare concepts. The common binary code space is built on the joint learning of the visual representations attended by language, the semantic-invariant structure construction and the label distribution correction. Extensive experiments demonstrate the superiority of our method.
\end{abstract}

\maketitle

\section{Introduction}
In this big data era, large volume and high-dimensional multimedia data is ubiquitous in social networks and search engines. This leads to the major challenge of how to efficiently retrieve information from the large-scale database \cite{wang2018survey}. To guarantee retrieval efficiency and quality, approximate nearest neighbour (ANN) search has attracted increasing attention in recent years. Parallel to the traditional indexing methods, hashing is one of the most advantaged methods in existing ANN methods, as it transforms high dimensional multimedia data into compact binary codes and enables efficient xor operations to accelerate calculation in Hamming space. In this paper, we will focus on learning to hash methods which build upon data-dependent binary encoding schemes for efficient image retrieval, which have demonstrated superior performance over data-independent hashing methods, e.g. LSH~\cite{lsh}.

\begin{figure}[!t]
\includegraphics[height=4.4cm]{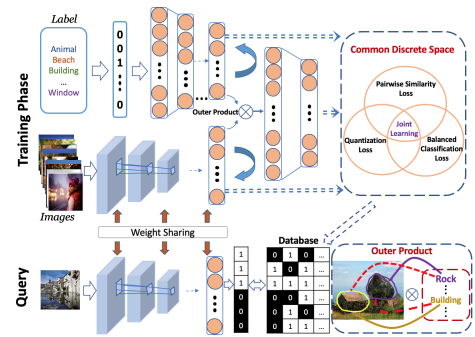}
\caption{\small The proposed deep collaborative discrete hashing framework. A dual-stream network consists of feature embedding network and label encoding network. We sketch the strength of the outer product in the bottom right subfigure. 
} \label{framework}
\end{figure}

Generally, learning to hash methods can be divided into unsupervised and supervised groups. Compared with unsupervised methods~\cite{ZZhang2018,ITQ}, supervised methods~\cite{sdh, cosdish, lfh, ZZhangAAAI, ZZhang2019} can yield better performance with the support of label supervision. With the rapid development of deep neural network, deep hashing methods~\cite{DHN, DQN, DPSH, dvsh, hashing, hashing1, hashing2} have demonstrated superior performance over non-deep hashing methods and achieved state-of-the-art results on public benchmarks. 

However, among mainstream deep hashing frameworks, human-annotated labels purely supervise the distribution alignment of hash code embedding, yet fail to trigger context-aware visual representation learning, let alone optimal binary codes generation. Moreover, the correlations between features and semantics are not well-explored to generate semantic consistent binary codes. Furthermore, existing supervised methods are vulnerable to the imbalanced distribution of semantic labels. Models tend to grasp the frequently appeared concepts in the training data and disregard the infrequent ones, which highly restricts the expression capacity of hash codes. Hence, existing deep hashing methods may fail to generate optimal binary codes for efficient image retrieval.

In this paper, we propose a novel Deep Collaborative Discrete Hashing (DCDH) method, which constructs a discriminative common discrete space via dual-stream learning, as illustrated in Figure \ref{framework}. The main idea of the proposed framework is to construct a semantic invariant space, via bridging the gap between visual space and semantic space. Specifically, (1) We develop a bilinear representation learning framework, which significantly fuses and strengthens visual-semantic correlations to learn context-aware binary codes. (2) We employ outer product on visual features and label embeddings to generate more expressive representations rather than element-wise product or plain concatenation. To the best of our knowledge, this is one of the first attempts to utilize the outer product to capture pairwise correlations between heterogeneous spaces. (3) We seamlessly integrate our framework with the focal loss to enhance the discriminant of generated binary codes and mitigate the class-imbalance problem by reducing weights on the well classified concepts and increasing weights on rare concepts. (4) Extensive experiments conducted on benchmark datasets demonstrate that DCDH is capable to generate more discriminative and informative binary codes and yield state-of-the-art performance.

\section{The proposed Approach}\label{method}
\subsection{Problem Formulation}
Given a set of $n$ images $ \mathcal{D} = \{\bm x_i, \bm l_i\}_{i=1}^n$, where $\bm x_i$ and $\bm l_i \in \{0,1\}^{c}$ are the $i$-th image and corresponding one-hot label vector, respectively. Deep hashing aims to 
encode data $X \in \Re^{n\times d}$ as $k$-bits binary codes $B\in \Re^{n\times k}$. In our method, we mainly focus on the pairwise similarity-preserving hashing. In particular, we construct the similarity information based on ground-truth label. If two images $i$ and $j$ share at least one common label, we define $i$ and $j$ are semantically similar and $S_{ij}=1$, otherwise $S_{ij}=-1$ indicating dissimilar.

\subsection{Deep Visual Embedding Network}
The purpose of the visual embedding network is to generate discriminative hash codes such that similar pairs can be distinguished from dissimilar pairs. Specifically, Hamming distance between $b_i$ and $b_j$ should be minimized when $S_{ij}$ = 1, while maximized when $S_{ij}$ = -1. To preserve the pairwise similarities \cite{ksh}, our work adopts smooth $L_{2}$ loss defined on the inner product between binary codes as:
\begin{equation}\label{j1}
\min_{b_i,b_j} \mathcal{L}_{1} = \sum_{i=1}^n\sum_{j=1}^n\| b_i^Tb_j - k s_{ij}\|_F^2  ~s.t.~b_i,b_j\in\{-1,1\}^k,
\end{equation}
However, it is difficult to generate the \textit{discrete} outputs. We can set $b_i = sgn(F_v(x_i;\theta_v))$, where $\theta_v$ denotes the parameters of deep visual embedding network. 

\begin{equation}\label{j2}
\begin{split}
\min_{\theta,B} \mathcal{L}_{1} =\| sgn(F_v(x_i;\theta_v))^TB - \rho{S}\|_F^2~s.t.~B\in\{-1,1\}^{n\times k},
\end{split}
\end{equation}
where $S\in\{1,-1\}^{n\times n}$ is the binary similarity matrix. In this paper, we designed an end-to-end feature learning network which extends the pretrained AlexNet~\cite{alexnet} model for discriminative visual embedding learning. Based on this backbone network, we replace the final classifier layer with a fully connected layer to transform the convolutional feature maps into the $k$-dimensional continuous codes $U$. Subsequently, we apply hyperbolic tangent (tanh) as the activation function to approximate non-differential signum (sgn) function, \textit{i.e.,} $b_i = sgn(u_i)$.  To control the quantization error and bridge the gap between the binary codes and its relaxation, we add an extra penalty term to keep $U_{v}$ and $B$ as close as possible. We adopt the following matrix-form loss function to facilitate the network back-propagate the gradient to $\theta_v$. Hence, the problem in (\ref{j2}) is transformed into the following problem:
\begin{align} \label{imagenet}
&\min_{B,\theta_v} \mathcal{L}_{1} = \| {U_v^TB} - \rho{S}\|_F^2 + \alpha \|  B -  U_v\|_F^2\\&s.t.~B\in\{-1,1\}^{n\times k}, U_{v} = tanh(F_v(x_i;\theta_v)),\nonumber
\end{align}
where $\alpha$ is a weighting parameter.

\subsection{Deep Class Encoding Network}
In pairwise-preserving hashing methods, labels are always exploited as similarity measurement between data points by applying the element-wise inner product. However, solely using similarity matrix to supervise hash codes learning inevitably results in severe information loss and thus highly restricts the expression ability of generated hash codes, especially in multi-label cases. To be more specific, one image annotated by multiple labels (such as 'ocean', 'beach' and 'water') contains underlying semantic connections in concepts, while single class vector may hinder the conceptual bridge at a fine-grained level. The purpose of the label encoding network is to capture the original semantic information and preserve them in $k$-dimensional flexible continuous space. Similarly, the loss function of the label encoding network can be defined as:
\begin{equation}\label{labelnet}
\begin{split}
\min_{B,\theta_l} \mathcal{L}_{2} &= \| {\text{tanh}(U_l)^TB} - \rho{S}\|_F^2 + \alpha \| B -  \text{tanh}(U_l)\|_F^2 \\
s.t.\quad U_l &= F_l(L;\theta_l),
\end{split}
\end{equation}
where $F_l(\cdot;\theta_l)$ denotes the label encoding network parameterized by $\theta_l$. By providing with complementary views of semantics, the label encoding network potentially guides the visual embedding network to learn beneficial context-aware representations.

\subsection{Semantic Invariant Structure Construction}
To disentangle the relationships between the abstract concepts and the visual features, we apply the outer product to fuse visual and label embeddings. Being distinct from the conventional element-wise product or plain concatenation, the applied outer product allows high-level label encoding and low-level visual feature embeddings to interactively influence each other. In this way, we can capture the pairwise correlations between the feature of an image and its corresponding label, enabling discovery of the common latent attributes. By applying the outer product, we first obtain the pairwise interaction between label and image features. After the training procedure, the semantic information is well separated by the related region in the image. The latent vector is obtained by reshaping the pairwise correlation matrix to a vector, which can project to discrete space to generate hash codes. The generated codes are more discriminative since the outer product operator ensures the bits truly reflect regions in the images to the corresponding semantic information. To construct the semantic invariant structure, the objective function can be formulated as:
\begin{equation}
\begin{split}
&\min_{B,\theta} \mathcal{L}_{3} = \| {\text{tanh}(U)^TB} - \rho{S}\|_F^2 + \alpha \|  B -  \text{tanh}(U)\|_F^2\\&
s.t.\quad U = F( U_v \otimes U_l;\theta),
\end{split}
\end{equation}
where $\otimes$ denotes the outer product and $F(\cdot;\theta)$ denotes the fusion network parameterized by $\theta$.

Furthermore, we introduce the focal loss \cite{focal} to mitigate the side effect from class imbalance, and the objective function is:
\begin{equation}\label{invariant}
\begin{split}
\min_{B,\theta,\theta_c} \mathcal{L}_{3} &= \| {\text{tanh}(U)^TB} - \rho{S}\|_F^2 + \alpha \|  B -  \text{tanh}(U)\|_F^2\\ &+ \sum_{i=1}^n\sum_{t=1}^c-(1-\hat{p_{i,t}})^{\gamma}log(p_{i,t})
\\
s.t.\quad p_{i} &= \text{softmax}(F_c(u_i;\theta_c)),
\end{split}
\end{equation}
where $\gamma$ is the hyper-parameter. The $F_c(\cdot;\theta_c)$ denotes the classification layer parameterized by $\theta_c$, and $p_{i,t}$ is the estimated probability of the $i$-th sample for the $t$-th class. The adaptive factor $\hat{p_{i,t}}$ is:
\begin{equation}
\hat{p_{i,t}} = \begin{cases}
p_{i,t}, & \text{if $l_{i,t}=1$};\\
1 - p_{i,t}, & \text{otherwise}.
\end{cases}
\end{equation}

\subsection{Collaborative Learning}
To learn context-aware and discriminative hash codes, we adopt the joint learning framework consisting of the visual embedding network, the label encoding network and the semantic-invariant structure construction, and we have:
\begin{equation} \label{collaborative}
\min_{B,\theta_v,\theta_l,\theta_c,\theta}\mathcal{L} = \mathcal{L}_1 + {\lambda} \mathcal{L}_2 + {\mu} \mathcal{L}_3,
\end{equation}
where $\lambda$ and $\mu$ are coefficients to weight the importance of different terms. ${\mathcal{L}_1}$, ${\mathcal{L}_2}$ and ${\mathcal{L}_3}$ denote the visual embedding net loss, the label encoding net loss and semantic invariant structure construction loss, respectively.

\subsection{Optimization}
The proposed DCDH needs to jointly optimize Eqn. (\ref{imagenet}), (\ref{labelnet}), and (\ref{invariant}). Due to similar forms, we only illustrate one detailed optimization on Eqn. (\ref{invariant}), and the rest equations can be solved similarly. Specifically, we adopt the iterative alternating optimization manner, that is, we update one variable with others fixed.
\\ \textbf{Learning $\theta_v$}: The network parameters ${\theta_v}$ can be optimized via standard back-propagation algorithm by automatic differentiation techniques in Pytorch \cite{pytorch}.
\\ \textbf{Learning $B$}: We aim to optimize $\widetilde{B}$ with all hyper-parameters fixed, and we rewrite the Eqn. (\ref{invariant}) as follows:
\begin{equation}\label{step1}
\begin{split}
\min_{B} \mathcal{L}_{1} &=  \| {\text{tanh}(U)^TB}\|_F^2 -  2\rho\Tr(B^T\text{tanh}(U)S)\\
&- 2{\alpha}\Tr(B^T\text{tanh}(U)) + const,
\end{split}
\end{equation}
where $\Tr(\cdot)$ is the trace norm. Since focal loss is independent to $\widetilde{B}$ update, we can consider focal loss as constant when learning $\widetilde{B}$.

It is challenging to directly optimize $\widetilde{B}$ due to its discrete constraint. Inspired by~\cite{sdh}, we learn binary codes $B$ by the DCC strategy, in which non-differential variable can be solved in a bit-by-bit manner. Therefore, problem (\ref{step1}) can be reformulated to minimize
\begin{align}\label{optim}
&\| {\text{tanh}(U)^TB}\|_F^2 - 2\Big(\rho\Tr({B}^T\text{tanh}(U)S) + \gamma Tr(B^T\text{tanh}(U) \Big) + const \nonumber \\
&= \Tr\Big(2(u^T(U^{\prime})^TB^\prime - q^T)z\Big) + const\quad s.t. B\in \{-1,1\},
\end{align}
where $Q = \rho\text{tanh}(U)S + \gamma \text{tanh}(U)$, and $\Bar{U} = \text{tanh}(U)$. $u^T$ is the row of $\Bar{U}$, $U^\prime$ denotes the matrix of $\Bar{U}$ exclude $u$. $z^T$ is the row of $B$, $B^\prime$ denotes the matrix of $B$ exclude $z$. $q^T$ is the row of $Q$, $Q^\prime$ denotes the matrix of $Q$ exclude $q$
Eventually, we can get the following optimal solution of problem (\ref{optim}) that can be used to update $z$:
\begin{equation} \label{optimize}
z = sgn(u^T(U^{\prime})^TB^\prime - q^T).
\end{equation}
The network parameters can be efficiently optimized through standard back propagation algorithm by using automatic differentiation techniques by PyTorch~\cite{pytorch}.

\subsection{Out of Sample Extension}
Based on the proposed optimization method, we can obtain the optimal binary codes for all the training data and the optimized visual embedding learning network, \textit{i.e.,} $F_v(x_i;\theta_v)$. Our learning framework can easily generate the binary codes of a new query ${x_q}$ by using the visual network followed by the signum function:
\begin{equation}
\begin{split}
    b_q = \phi(x_q; {\theta_v})& = \text{sgn}(F_v(x_q;\theta_v))
\end{split}
\end{equation}

\begin{table}
\begin{center}
\caption{\small The averaged retrieval MAP comparison on NUS-WIDE, and MIRFlickr. The best performance are shown in boldface.} \label{Table_1}\vspace{-0.3cm}
\resizebox{0.47\textwidth}{!}{
\begin{tabular}{|c|cccc|cccc|}
\hline
\multirow{2}{*}{\textbf{Methods}}&
\multicolumn{4}{c|}{\textbf{NUS-WIDE}}&
\multicolumn{4}{c|}{\textbf{MIRFlickr}}\\
\cline{2-9}
&12-bits & 24-bits & 32-bits & 48-bits
& 12-bits & 24-bits & 32-bits & 48-bits\\
\hline
LSH&0.3942 &0.4049 & 0.4305& 0.4331
&0.5456&0.5501&0.5460&0.5523\\
ITQ&0.5435 &0.5544 &0.5536 &0.5560
&0.6243 &0.6305 &0.6318 & 0.6359
\\
KSH&0.5701 &0.5735 & 0.5797& 0.5788  
 &0.6135 & 0.6144 &0.6213 & 0.6176 
\\
SDH&0.6769 & 0.6914 & 0.6981& 0.7052
&0.8018 &0.8258 & 0.8267&0.8387
\\
LFH&0.7152 & 0.7446 & 0.7512 & 0.7722
&0.8258 &0.8364 & 0.8281& 0.8573
\\
COSDISH&0.7398 & 0.7678 & 0.7819& 0.7888
&0.7736 & 0.7973& 0.8589&0.8693
\\
DHN& 0.7719 & 0.8013 & 0.8051 & 0.8146
& 0.8092 & 0.8283 & 0.8290 & 0.8411
\\
DVSQ& 0.7856 & 0.7924 & 0.7976 & 0.8019
&0.8112 & 0.8263 & 0.8288 & 0.8341
\\
DPSH& 0.7941 & 0.8249 & 0.8351 & 0.8442 
& 0.6857 & 0.7058 & 0.7140 & 0.7182
\\
DTQ& 0.7951 & 0.7993 &0.8012 & 0.8024
& 0.8098& 0.8274 & 0.8456 & 0.8511
\\
\hline
\textbf{DCDH (ours)}&\textbf{ 0.8223} & \textbf{0.8526} & \textbf{0.8712} & \textbf{0.8974}
& \textbf{0.8758} & \textbf{0.8970} & \textbf{0.9059} & \textbf{0.9079}\\
\hline
\end{tabular}}
\end{center}
\end{table}

\section{Experiments}\label{experiment}
We conduct extensive experiments to evaluate our method against several state-of-the-art hashing methods on NUS-WIDE and MIRFlickr. NUS-WIDE contains 269,648 images with 81 tags. Following \cite{DPSH}, we select a subset of 195,834 images that are included in the 21 most frequent classes. MIRFlickr contains 25,000 images from Flickr website, in which each image is tagged by at least one of 38 concepts. Following evaluation splits in \cite{sdh, cosdish}, we randomly sample 2,100 and 1700 images as query sets for NUS-WIDE and MIRFlickr, respectively, and the rest are utilized for training. 

We compare our DCDH with 10 state-of-the-art hashing methods, which include 4 non-deep hashing methods (i.e. KSH~\cite{ksh}, SDH~\cite{sdh}, COSDISH~\cite{cosdish}, LFH~\cite{lfh}), 4 deep hashing methods (i.e. DPSH~\cite{DPSH}, DHN~\cite{DHN}, DVSQ~\cite{dvsh}, DTQ~\cite{dtq}), 1 unsupervised method (i.e. ITQ~\cite{ITQ}) and 1 data independent method (i.e. LSH~\cite{lsh}). For fair comparison, we employ 4096-dim deep features extracted from AlexNet~\cite{alexnet} for non-deep methods. Two evalution metrics, \textit{i.e.,} Mean Average Precision (MAP), and Precision@top K, are used for performance comparison.

\subsection{Results}
\begin{figure}[t]
\begin{tabular}{cc}
\includegraphics[width=0.24\textwidth]{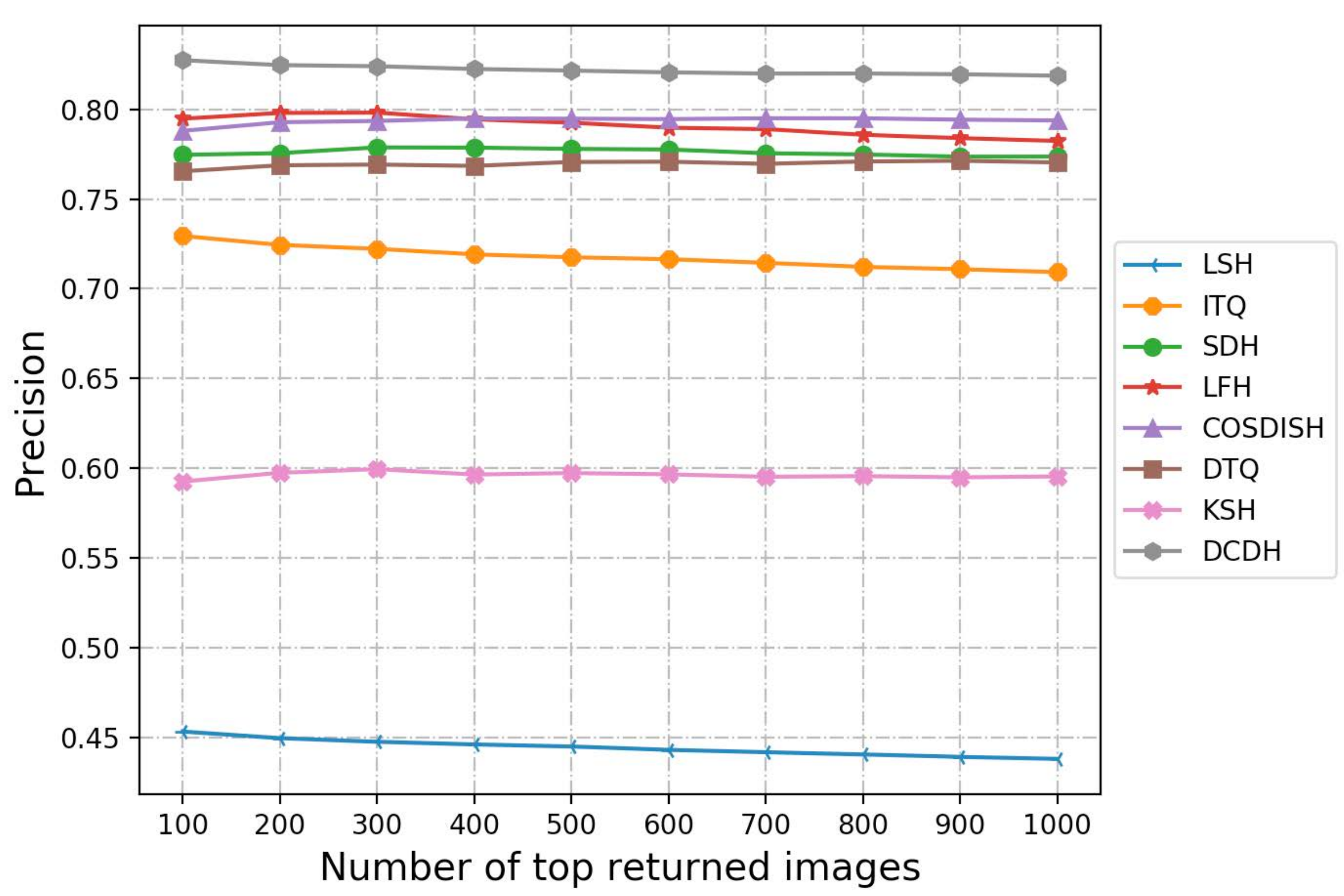}&
\includegraphics[width=0.24\textwidth]{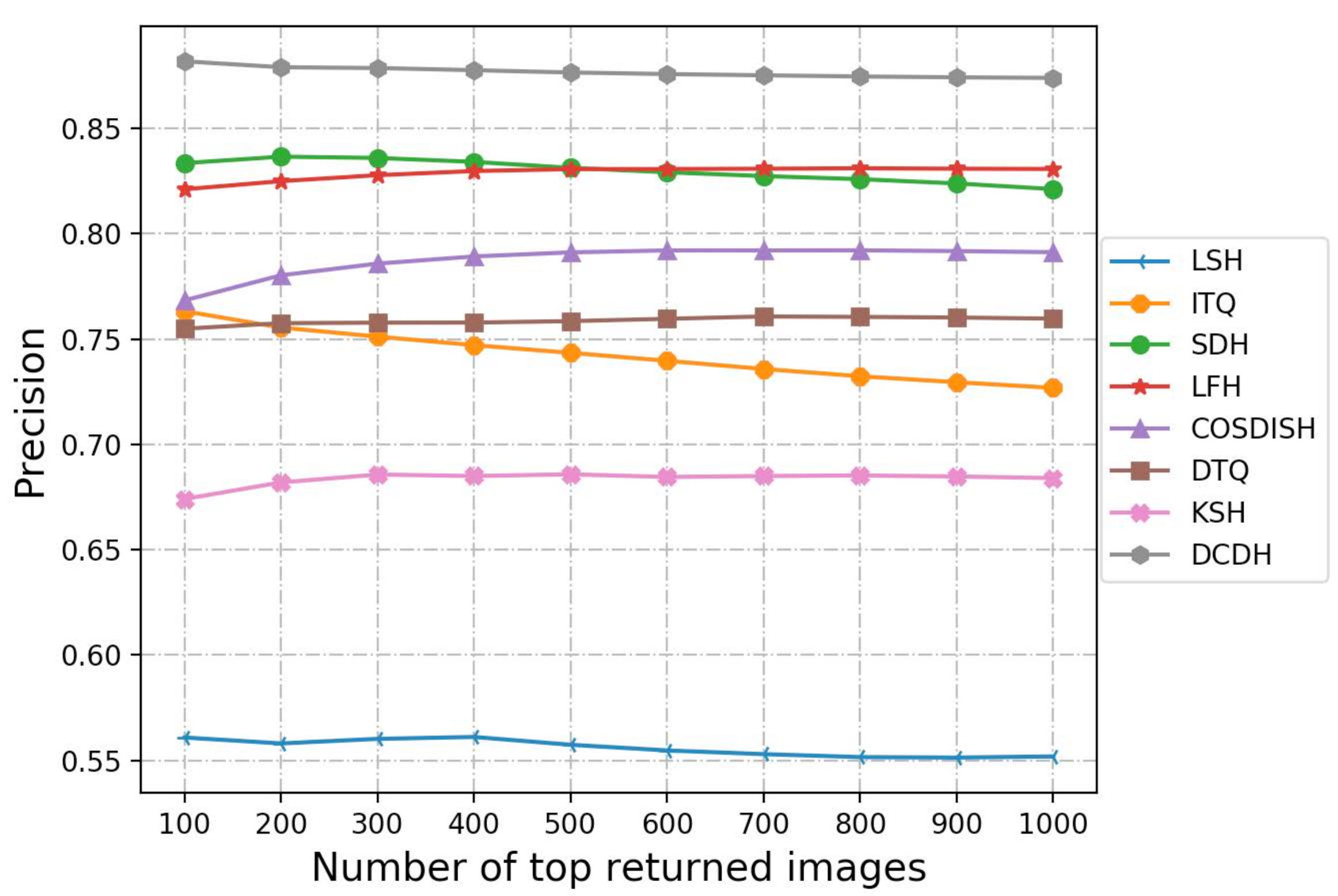}\\
\small(a) 12 bits@NUS-WIDE  &
\small(b) 12 bits@MIRFLickr\\
\end{tabular}
\caption{\small The precision curves of the top-N returned samples on the NUS-WIDE in (a) and MIRFlickr in (b).}\label{pcurve} 
\end{figure}

The MAP results of different methods on NUS-WIDE, and MIRFlickr are reported in Table \ref{Table_1}. (1) Generally, taking advantage of semantic information, supervised methods can achieve better retrieval performance than unsupervised methods, while ITQ can obtain competitive results. (2) Deep hashing methods can outperform shallow methods in most cases, since deep hashing methods benefit from learning discriminative representations and non-linear hash functions. (3) From MAP results in Table \ref{Table_1} and percision@top K curves in Figure \ref{pcurve}, we can observe DCDH outperforms other comparison methods by a large margin. Our proposed method always produces the best performance on both of the benchmarks, which emphasizes the importance of semantic invariant structure construction and excavating the underlying semantic correlation.

\begin{table}
\begin{center}
\caption{\small Ablation study of our DCDH method.}\label{t2}
\resizebox{0.32\textwidth}{!}{
\begin{tabular}{|c|cc|cc|}
\hline
\multirow{2}{*}{\textbf{Methods}}&
\multicolumn{2}{c|}{\textbf{NUS-WIDE}}&
\multicolumn{2}{c|}{\textbf{MIRFlickr}}\\
\cline{2-5}
 & 12-bits & 48-bits & 12-bits & 48-bits\\
\hline
DCDH-V& 0.7477 & 0.8153 & 0.7842 & 0.8681\\
DCDH-S& 0.7677 & 0.8523  & 0.8146 & 0.8875\\
\hline
DCDH&\textbf{0.8223} &\textbf{0.8974} &\textbf{0.8758} &\textbf{0.9079}\\
\hline
\end{tabular}}
\end{center}
\end{table}

\subsection{Ablation Study}
We investigate two variants of DCDH: 1) DCDH-V utilizes the visual feature embedding net solely to generate hash codes. 2) DCDH-S leaves alone the semantic encoding and visual feature embedding to generate hash codes. We report the results of DCDH variants in Table \ref{t2} with 12 bits hash codes and 48 bits hash codes on NUS-WIDE, and MIRFlickr. Compared with full model DCDH, we observe that both DCDH-S and DCDH-V incur a descendant in MAP. DCDH-S can achieve better performance than DCDH-V after employing the supervision of encoded labels. The result further reveals that the importance of mining the semantic correlation between semantic information and local visual features.

\section{Conclusion}\label{conclusion}
In this paper, we proposed a novel deep supervised hashing framework, which collaboratively explores the visual feature representation learning, semantic invariant structure construction, and label distribution correction. A discriminative common discrete Hamming space was constructed by concurrently considering the shared and model-specific semantic information from visual features and context annotations. Moreover, the class imbalance problem was addressed to leverage frequent and rare concepts. Extensive experimental results demonstrate the superiority of the proposed joint learning framework. 

\begin{acks}
This work is supported by ARC discovery project DP190102353.
\end{acks}

\bibliographystyle{ACM-Reference-Format}
\bibliography{sigir}

\end{document}